# High Performance Graphene Integrated Photonics Platform Enabled by Gold-assisted Transfer

XIAOXUAN WU,[1,†] ZHENGYI CAO,[2,†] TIANXIANG ZHAO,[1] YUN WU,[2,*]
ZHONGHUI LI,[2] SPYROS DOUKAS,[3] ELEFTERIOS LIDORIKIS,[3] YU XUE,[4]
LIU LIU,[4,5] OMID GHAEBI,[6] GIANCARLO SOAVI,[6,7] JUNPENG LV,[1,*]
ZHENHUA NI,[1,*] AND JUNJIA WANG[1,*]

[1]*National Research Center for Optical Sensors/Communications Integrated Networks, School of Electronic Science and Engineering, Southeast University, Nanjing 210096, China*
[2]*CETC Key Laboratory of Carbon-based Electronics, Nanjing Electronic Devices Institute, Nanjing 210016, China*
[3]*Department of Material Science and Engineering, University of Ioannina, GR 45110 Ioannina, Greece*
[4]*State Key Laboratory for Modern Optical Instrumentation, College of Optical Science and Engineering, International Research Center for Advanced Photonics, Zhejiang University, Hangzhou 310058, China*
[5]*Jiaxing Key Laboratory of Photonic Sensing & Intelligent Imaging, Intelligent Optics & Photonics Research Center, Jiaxing Research Institute, Zhejiang University, Jiaxing 314000, China*
[6]*Institute of Solid State Physics, Friedrich Schiller University Jena, 07743 Jena, Germany*
[7]*Abbe Center of Photonics, Friedrich Schiller University Jena, 07745 Jena, Germany*
[†]*These authors contributed equally to this work.*
*\*wuyun012@126.com*
*\*phyljp@seu.edu.cn*
*\*zhni@seu.edu.cn*
*\*junjia_wang@seu.edu.cn*

**Abstract:** Graphene is promising for nanoscale, efficient, ultra-fast photo- and opto-electronic devices because of its remarkable electrical and optical properties, such as fast electron relaxation and heat dissipation. Here, we realize high-performance graphene integrated photonics platform enabled by gold-assisted transfer. Thanks to our optimized transfer technique, we fabricate and demonstrate (1) a microscale thermo-optic modulator with a tuning efficiency of 0.037 nm/mW and high heating performance of 67.4 K·µm$^3$·mW$^{-1}$ on a small active area of 7.54 µm$^2$, (2) a graphene electro-absorption modulator featuring an high modulation bandwidth up to 26.8 GHz and a high-speed data rate reaching 48 Gb/s, and (3) a graphene Mach-Zehnder interferometer modulator with a high normalized modulation efficiency of 0.027 dB·V$^{-1}$·µm$^{-1}$.

Our graphene integrated photonics platform has far superior performances compared to state of the art in terms of efficiency, low process complexity, and compact device footage. Thus, our approach and results provide the background for the realization of high-performance integrated photonic circuits with CMOS compatibility.

## 1. Introduction

Silicon on insulator (SOI) technology has become an attractive platform for high-efficiency photonic and opto-electronic applications due to its complementary metal-oxide-semiconductor (CMOS) compatibility, low-complexity fabrication processes, and low-cost manufacturing techniques [1]. However, the efficiency and speed of SOI-based optical modulators are limited by the intrinsic properties of silicon. In particular, being a centrosymmetric crystal, modulation of absorption and the phase of the incident electromagnetic field *via* an external electric field (Pockels effect or Kerr effect) in silicon is challenging. Possible routes to indirectly modify light-matter interactions in silicon include doping and structural modifications [2]. Furthermore, the plasma dispersion effect in silicon facilitates electro-optical modulation in waveguides [3]. However, the limited bandwidth in silicon electro-optical modulators poses challenges for practical applications, such as communication networks, requiring higher information capacity and faster processing speed. Therefore, the development of hybrid devices on the SOI platform is highly desirable and has considerable technological relevance.

In this context, graphene is arguably the most promising candidate for integrated opto-electronics [4]. In the last two decades, graphene has been subject of intense research which led to the realization of micro-scale optoelectronic devices [5–7] with advanced functionalities, such as linear and nonlinear optical modulators [8–11], photodetectors [12], light emitters [13], gas sensors and gates for logic operations. Such applications were made possible thanks to the combination of high carrier mobility [14] (200,000 $cm^2 \cdot V^{-1} \cdot s^{-1}$), ultrafast hot electron dynamics [15], large thermal conductivity [16] (5300 W/mK) and strong light-matter interactions in the linear and nonlinear regimes [17], which all together allow opto-electronic operations at high speed and with high efficiency.

There are mainly two mechanisms that can be exploited to realize graphene-based light modulators: the thermo-optic (TO) effect and the electro-absorption (EA) effect. Graphene heaters have proven to be highly efficient in TO modulators, thanks to their low optical absorption [18] and high thermal conductivity. Taking advantage of the TO effect, graphene has been incorporated into various waveguide structures such as Mach-Zehnder interferometers [19,20] (MZI), micro-ring resonators [21,22], and photonic crystal nanocavities [23,24] to achieve energy-efficient light modulation. On the other hand, graphene EA modulators can convert electric signals into a modulation of the light intensity, enabling effective amplitude modulation. Various graphene EA modulators have been developed, including graphene-silicon [25], suspended graphene [26], double-layer graphene [27–29] (DLG), and graphene-based hybrid plasmonics [30], each exhibiting pro and cons. To date, graphene modulators based on the SOI platform are still regarded as one of the most promising candidates for large scale photonic systems [31].

Although a lot of work has been done, there are still challenges to achieving practical solutions in many key applications, and the exploitation of graphene latent capabilities remains an ongoing endeavor. For instance, if we consider the bandwidth of graphene modulators, a notable disjunction persists between the maximum experimental bandwidth [32,33] and the corresponding theoretical projections (~500 GHz [34]). Moreover, the deployment of graphene modulators is presently limited to laboratory-based investigation, predominantly due to the intricacies of their manufacturing processes and the low production yield. In this context, the exploration of novel graphene integrated photonics platforms is highly desirable.

In this work, we propose a graphene SOI integrated photonics platform enabled by the gold-assisted transfer method, which is expected to provide wafer-scale implementation and high-performance devices. Two types of modulators are fabricated to confirm the advantages of our approach. We first demonstrated a TO modulator based on a silicon micro-ring structure (Q ~67000) with a graphene heater integrated on it. Due to efficient thermal tunability of the refractive index *via* an external electric field, a heating efficiency as high as 67.4 $K \cdot \mu m^3 \cdot mW^{-1}$ have been demonstrated over a small active area of 7.54 $\mu m^2$. At the same time, the device possesses high modulation speed with 10−90% rise and fall response times of 4.8 and 5.5 μs, respectively. These results are significantly faster compared to metal TO modulators, which typically exhibit response times of a few milliseconds [35,36]. We further realized a graphene EA modulator based on the same architecture of the TO modulator, where we simply replaced the $AlO_x$ with a $HfO_2$ dielectric layer. This device demonstrates a high 3 dB bandwidth up to 26.8 GHz and a high data rate reaching 48 Gb/s with a modulation efficiency of 1.3 dB/V. The MZI device exhibits a high modulation depth up to 13.3 dB with a bias voltage of only 5 V and a high modulation efficiency up to 5.3 dB/V, which is the highest efficiency among graphene EA modulators to the best of our knowledge. The data rate is measured to be 32 Gb/s with a 3 dB bandwidth of 7.4 GHz. The proposed platform and devices provide a feasible solution to support large-scale high-performance photonic systems and paves the way for future optical communication and sensing applications.

## 2. Results and discussion

### 2.1 Gold-assisted transfer method

The conventional wet-transfer method, which is currently the standard approach for the fabrication of opto-electronic devices, employs polymethyl-methacrylate polymer (PMMA) or other organic compounds as the supporting layer [37,38]. There are some remarkable improvements of this method but the remove of supporting layer is generally necessary with organic procedure such as acetone steam [39–41]. However, due to the strong chemical adsorption between graphene and organic compounds, complete removal of organic residues is usually impossible [42], leading to poor metal-graphene contact as well as performance degradation. Another approach for the fabrication of graphene based opto-electronic devices is dry-transfer method but with low yield and incapability of wafer-scale fabrication. For both of our devices, we employed the gold-assisted transfer process, as shown schematically in Fig. 1. Compared to other graphene transfer method, the gold-assisted method, which adopts gold films instead of PMMA as the supporting layer, provides simplified fabrication and low contact resistances [43,44]. The process begins by depositing a gold supporting layer onto a commercially available single-layer graphene (SLG) sample grown on a copper foil. Subsequently, the copper substrate undergoes etching by floating it on an ammonium persulfate solution for a duration of 4 hours. Once the copper etching process is completed, the gold-SLG film is carefully transferred into a beaker containing deionized (DI) water and then carefully lifted onto the target substrate. After ensuring the complete drying of the transferred film, any excess gold is removed using an aqueous acid solution, while the excess graphene films are etched through oxygen plasma. The graphene layers in our devices exhibit a surface resistance of approximately 700 Ω/□ and a contact resistance of approximately 0.3 Ω·mm (see Supplementary Note 1 for more details).

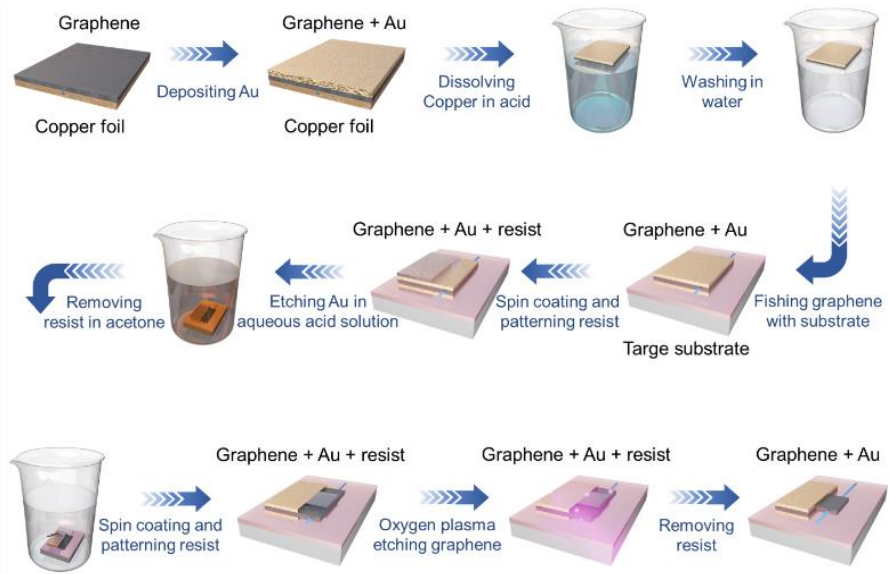

**Fig. 1. Gold-assisted transfer method.** The process begins with depositing Au supporting layer on commercial SLG sample grown on copper foil. Then the copper substrate is completely etched by placing it in ammonium persulfate solution for 4 hours. Once the copper etching is complete, the Au-SLG film is transferred in a beaker with DI water and then lifted by the target substrate. When completely dry, the excess Au and graphene films is removed using aqueous acid solution and oxygen plasma etching, respectively.

### 2.2 Graphene TO modulator

Figure 2a shows a sketch of the graphene TO micro-ring modulator. The resonator consists of a micro-ring with a radius of 30 μm and a gap of 200 nm coupled with two bus waveguides. Four grating couplers are used for coupling the signal light in and out of the chip. This configuration is highly sensitive to small change of the mode effective refractive index ($n_{eff}$). Figure 2b shows the cross-section view of the device. As depicted in Fig. 2b, a buried waveguide with a cross-section of 220 nm×480 nm was used to guide single-mode transverse electric (TE) light. The simulated electric field distribution of the guided mode is shown in Fig. 2c. On the top of the waveguide, 10 nm of benzocyclobutene (BCB) layer was spin-coated to enhance the adhesion ability of the substrate with graphene sheets and thus facilitate the transfer of high-quality SLG. Further, a SLG/AlO$_x$/SLG structure consisting of two single layer graphene with a 20 nm AlO$_x$ spacer is designed on top. The SLG/AlO$_x$/SLG structure overlaps above the resonator forming an arc along the ring with a central angle of 30 degrees and a width of ~1.5 μm above the resonator. Two gold electrodes were formed on both SLG sheets on each side of the waveguide to form a micro-heater. Ohmic heating is generated in the SLG *via* a potential difference between the gold electrodes and leads to a temperature increase on the silicon waveguide.

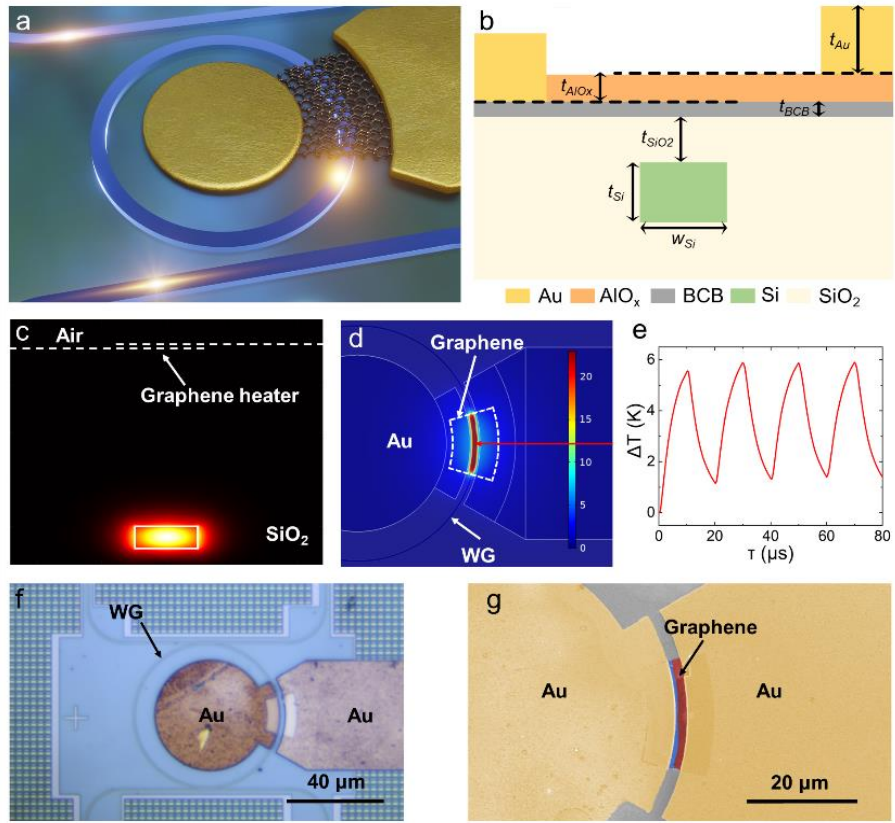

**Fig. 2. The graphene TO modulator. a,** 3D illustration of the graphene TO micro-ring modulator. **b,** Cross-section of the device. The black dashed line shows the graphene sheets. **c,** Waveguide TE-mode profile distribution simulated by finite-different time-domain simulation. **d,** Temperature rise distribution across the x-y plane for ~ 1 mW power dissipated in the proposed device. **e,** Si waveguide transient temperature rise at the point indicated with the red arrow. **f,** Optical microscope image of the device. Scale bar, 40 μm. **g,** False-color SEM image of the device. Scale bar, 20 μm.

To verify the thermal conduction performance in the heating process, the temperature distribution of the proposed device was simulated with the 3D finite element method using the COMSOL transient heat transfer module. Simulations were performed assuming thermal conductivities of ~ 5000 W·m$^{-1}$·K$^{-1}$ for graphene, 301 W·m$^{-1}$·K$^{-1}$ for Au contacts, 131 W·m$^{-1}$·K$^{-1}$ for Si waveguide and substrate, 1.12 for glass substrate and 30 W·m$^{-1}$·K$^{-1}$ for AlO$_x$. For the top boundary of the device, free convection to ambient air of $h_1$ = 15 W·m$^{-2}$·K$^{-1}$ was assumed. For the bottom part, we assumed loose contact with metal holder, with a heat transfer coefficient of $h_2$ = 150 W·m$^{-2}$·K$^{-1}$. Graphene layers are modeled as infinitesimally thin films. The heat dissipated in the device is distributed according to the resistance of each constituent element, *i.e.*, the overlapping and non-overlapping area of the graphene layers, the $AlO_x$ thin layer resistance and the contact resistances, considering their respective electrical resistance values (see Supplementary Note 4). The power is dissipated in a pulsed manner, assuming a total of 4 applied pulses, with 10 μs pulse width and 20 μs pulse period. All the geometric aspects of the simulations are set up according to the device used in our experiments. Figure 2d shows the temperature rise distribution across the x-y plane at its peak value, *i.e.*, at the center of the 4$^{th}$ pulse assuming a total of ~ 1 mW power deposited in the configuration, *i.e.*, 1 V applied bias. Figure 2e shows the transient temperature rise of the waveguide below the graphene layers. We extract the transient solution of temperature at the point indicated with the red arrow in the waveguide for different applied voltages. In the micro-heater system, the resistance within the $AlO_x$ layer serves to augment heating efficiency, while the exceptional thermal conductivity inherent in graphene facilitates rapid switching speeds.

The buried silicon waveguide was fabricated by typical lithography and dry-etching processes on a commercial SOI wafer with a 220 nm thick top silicon layer and a 2 μm thick $SiO_2$ buried oxide layer. Then the $SLG/AlO_x/SLG$ layers were assembled (see Methods for details) and carefully aligned with the micro-ring using the gold-assisted transfer techniques, as already discussed in details. Figure 2f shows a microscope image of the fabricated device, where the ring resonator and gold contacts can be seen clearly, while Fig. 2g shows the scanning electron microscope (SEM) images of the fabricated device, in which the top and bottom SLG (red and blue) overlap above the buried ring waveguide. The TE-polarized light is coupled from the fiber to the waveguide using the input grating coupler. By applying a bias voltage to the micro-heater to heat the waveguide underneath, leading to variations in the effective index of the transmission mode. Applying a bias voltage to the micro-heaters heats the underlying waveguide, resulting in the modulation in the effective index for the transmission mode. Consequently, these index variations result in a shift of the resonance peak in the transmission spectrum of the device. In turn, this allows for a large modulation of the light transmission at a fixed selected wavelength, such as ~1550 nm, finally leading to efficient TO modulation.

We first characterized the I-V properties of the graphene TO modulator and measured a total resistance of 1.16 kΩ (see Supplementary Note 4). The static electro-optic response of the device is measured using the experimental setup shown in the Methods and the free spectral range (FSR) and extinction ratio (ER) are approximately 3 nm and 19 dB, respectively (see Supplementary Note 2). In Fig. 3a, by increasing the external bias voltage applied to the graphene micro-heater, we observe a red shift of the resonance peak due to the TO effect, as discussed above. The wavelength shifts of the resonance peak relative to the voltages of 1, 2, 3, 4 and 5 V are 0.04, 0.2, 0.4, 0.7 and 1.1 nm, respectively. We numerically fit the experimental measurements with Lorentzian function in the black dot lines in Fig. 3a. Inset in Fig. 3b shows the change in the $n_{eff}$ with respect to the bias voltage. The estimated refractive indices at bias voltages of 0, 1, 2, 3, 4 and 5 V are 2.3933, 2.3941, 2.393, 2.4003, 2.4059 and 2.4131, respectively (see Supplementary Note 3). As the voltage is increased, localized heating occurs, leading to changes in the coupling state within the micro-ring resonator and then quality-factor of the device reduced gradually.

Figure 3b presents the shifts of the resonance peaks as a function of the electrical tuning power, from which we extract a tuning efficiency of 0.037 nm/mW by linear fitting, which can be improved by further reduced the thickness of the top oxide layer, thus bring the graphene closer to the waveguide. This translates into a power of 81 mW required to obtain one FSR shift. We define the effective tuning efficiency as the product of tuning power needed for one FSR shift and active area and thus obtain 610.74 mW·μm$^2$, considering the small active area of 7.54 μm$^2$ of our device.

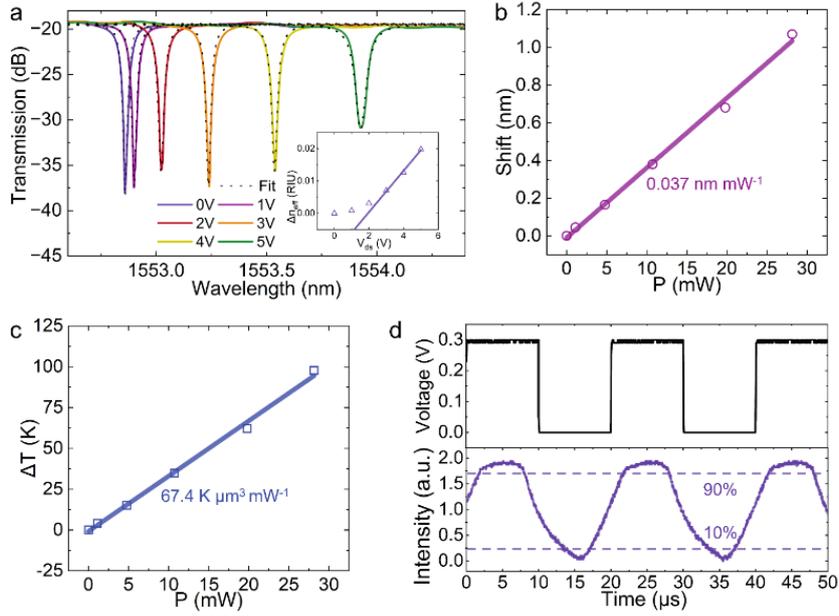

**Fig. 3. Electro-optic response of the graphene TO modulator. a,** Transmission spectra (colors) for various applied bias voltages (0-5 V). Inset shows the change in the effective refractive index in active area as a function of the bias voltages. **b,** Wavelength shift of the resonance peaks as a function of the input electrical power to graphene. **c,** Temperature change in the silicon resonator as a function of the input power to graphene. **d,** Normalized output intensity from the through port at a rectangular bias signal (1.85−2.15 V) at 50 kHz to graphene.

Next, we studied the heating performance of the graphene heater, which dissipates the electrical power into heat above the silicon micro-ring resonator, using a commercial finite-element solver. Taking into account the TO coefficient of silicon at 1550 nm ($\sim dn/dT = 1.8\times10^{-4}$ K$^{-1}$), the refractive index of silicon at different temperatures can be calculated, after which the corresponding $n_{eff}$ of the waveguide mode can be calculated. By fitting the extracted $n_{eff}$ with the measured one (inset in Fig. 3a), the produced temperature changes ($\Delta T$) in the silicon waveguide are determined as a function of the electrical power, as shown in Fig. 3c. The $\Delta T$ increases linearly with increasing electrical power, and the heating efficiency, which corresponds to the conversion efficiency from electrical power to Joule heating, can be estimated as 67.4 K·μm$^3$·mW$^{-1}$, considering the slope of the fitted line in Fig. 3b (3.39 K·mW$^{-1}$) and the volume of the resonator (19.9 μm$^3$) [19,21].

Further, we conducted real-time optical waveform measurements to quantify the dynamical electro-optical response of the graphene TO modulator. The normalized signal output from the through port is plotted in violet in Fig. 3d by driving the device with a square waveform electrical signal at 50 kHz (black curve). The modulated signal intensity was detected by a photodetector and recorded by an oscilloscope. The output signal responded to the input signal with 10−90% rise and fall times of 4.8 and 5.5 μs, respectively, which indicates that our device is faster than the previous metal TO micro-ring modulators reported.

In Table 1 we show a comprehensive comparison of the performances of our device with previously reported state-of-the-art TO modulators. Our graphene TO modulator has distinct advantages in terms of heating efficiency and normalized tuning efficiency compared to other graphene TO modulators (Refs 24 and 25). In addition, our experimental results confirm that graphene TO modulators exhibit higher efficiency and speed compared to TO modulators based on metal heaters (Refs 40 and 41).

**Table 1. Performances of the graphene TO modulators**

| Device structure | Rise time (μs) | Fall time (μs) | Tuning efficiency (nm/mW) | Heating efficiency (K·μm³·mW⁻¹) | Normalized tuning efficiency (mW·μm²) |
|---|---|---|---|---|---|
| Graphene micro-ring resonator(a) | 4.85 | 5.55 | 0.037 | 67.4 | 610.74 |
| Graphene micro-ring resonator [22] | 3.0 | 3.6 | 0.33 | 8.8 | 5104 |
| Graphene racetrack resonator [21] | 1.2 | 3.6 | 0.24 | 7.66 | 1484 |
| NiCr-Au micro-ring resonator [36] | 1500 | 2000 | 0.054 | N/A | N/A |
| Ti racetrack resonator [35] | 9 | 6 | 0.91 | N/A | N/A |

(a) This work;

*2.3 Graphene EA modulator*

As discussed, the response time of TO modulators is usually in the range of microseconds, ultimately limited by the heat dissipation process. EA modulators can achieve a significantly faster switching time (~ns) thanks to the high mobility of graphene. In our device, by simply replacing the conducting $AlO_x$ spacer with insulating $HfO_2$, we realized graphene EA modulators suitable for high speed applications. The sketch and cross-sectional view of the fabricated graphene EA modulator are displayed in Fig. 4a and 4c. A buried waveguide with a thickness of $t_{si}$ = 220 nm and a width of $w_{si}$ = 450 nm was employed to guide TE light around 1550 nm. In addition to the waveguide fabrication process, the silicon wafer undergoes chemical mechanical polishing to planarize the oxide layer and thus bring the waveguide closer to graphene. Then the $SLG/HfO_2/SLG$ structure was fabricated on top of the waveguide using the gold-assisted transfer method described in the previous section, which ensured the high quality of graphene along with low impurities and high quality interfaces. The thickness of the $HfO_2$ dielectric layer is 30 nm and the length of the $SLG/HfO_2/SLG$ stacks is set to be 180 μm. Figure 4d shows the simulated electrical field distribution of the guided mode, which illustrates that most of the optical power is confined in the waveguide core and only a small portion of the light (~1.1%) is evanescently coupled to the graphene layers. The structure configuration of our device can be clearly seen from the optical microscope image in Fig. 4b.

We measured the static electro-optic response of the EA modulator using the experimental setup described above. The modulation efficiency is determined to be 1.3 dB/V at 4 V (Fig. 4e). To confirm the high-speed capabilities of the graphene EA modulator, we conducted small signal measurements to determine its electro-optical bandwidth, as illustrated in Fig. 4f. A bias tee is used to combine a 3 V DC bias with a radio frequency (RF) signal provided by a photonic network analyzer (PNA) and then fed to the graphene EA modulator with a GS probe. The output light was amplified by an erbium-doped fiber amplifier and filtered by an optical tunable filter. Finally, it was transmitted to the PNA for detection and processing. The measured $S_{21}$ frequency response indicates that the 3 dB bandwidth of our device is ~26.8 GHz. The high-speed data transmission potential of the EA modulator is further verified by transmission measurements. A 2 V peak-to-peak $2^{15}-1$ pseudo-random binary sequence signal was generated by an arbitrary waveform generator with a DC bias of 3 V. Figure 4g depict the eye diagrams of non-return-to-zero (NRZ) signal generated at data rates of 48 Gb/s. Disregarding the parasitic capacitance, we achieved an energy efficiency of approximately 242 fJ/bit for the graphene EA modulator, which is calculated using the formula $C(V_{AC})^2/4$, with a measured capacitance (C) of 200 fF and a voltage swing ($V_{AC}$) of 2.2 V. This energy efficiency demonstrates the potential of our graphene EA modulator for high-speed and low-power applications.

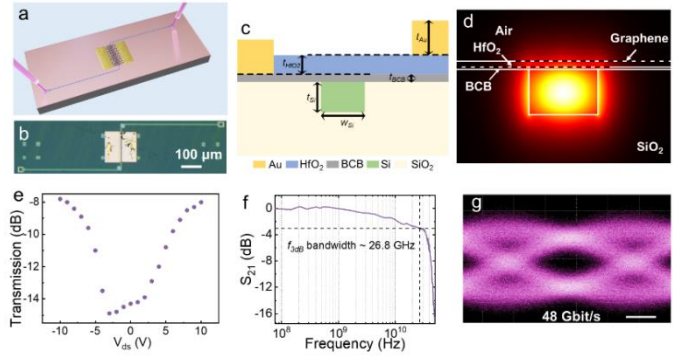

**Fig. 4. The graphene EA modulator. a,** 3D illustration of the graphene EA modulator. **b,** Optical microscope image of the device. Scale bar, 100 μm. **c,** Cross-section of the device. The black dashed line shows the graphene sheets. **d,** Waveguide TE-mode profile distribution. **e,** Corresponding transmission at 1550 nm as a function of applied bias voltages. **f,** Measured electro-optical $S_{21}$ frequency response of the EA modulator when the bias voltage is 3 V. **g,** NRZ eye diagrams generated at data rates of 48 Gb/s. The white scale bar corresponds to 5 ps.

Further, we fabricated a graphene MZI modulator with a path difference of 362 μm (Fig. 5a-b). The waveguide width of the MZI is 550 nm and the 200 nm long SLG/HfO$_2$/SLG structures are placed on both arms of MZI. We monitored the resonance peak around 1549.2 nm by applying a bias voltage on the electrodes of one arm. The transmission losses at 1549.2 nm for different voltages (0-5 V) are shown in Fig. 5c. The maximum loss is achieved when the voltage is 2 V. With a bias of only 5 V, the EA modulator exhibits a modulation depth as high as 13.3 dB. We calculate a modulation efficiency of 5.3 dB/V at 3 V, which corresponds to a multifold increase compared with previously reported graphene EA modulators [27,29,33,45]. Taking into account the length of the modulation structure (~200 μm), we calculate a normalized modulation efficiency of 0.027 dB·V$^{-1}$·μm$^{-1}$. The device exhibits a 3 dB bandwidth of approximately 7.4 GHz at a DC bias of 5 V, as evidenced by the measured $S_{21}$ frequency response in Fig. 5e. Figure 5f depict the eye diagrams of NRZ signal generated at data rates of 32 Gb/s. An eye diagram of PAM-4 is also shown in Fig. 5g with 16 Gbaud/s.

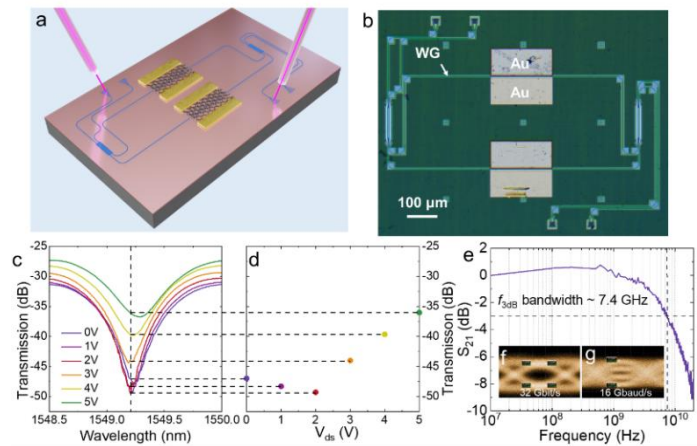

**Fig. 5. The graphene EA MZI modulator. a,** 3D illustration of the graphene EA MZI modulator. **b,** Optical microscope image of the device. Scale bar, 100 μm. **c,** Transmission spectra (colors) for various applied bias voltages (0-5 V) around 1549.2 nm. **d,** Corresponding transmission at 1549.2 nm as a function of applied bias voltages. **e,** Measured electro-optical $S_{21}$ frequency response of the EA modulator when the bias voltage is 5 V. **f,** NRZ eye diagrams generated at data rates of 32 Gb/s. **g,** PAM4 eye diagram generated at a data rate of 16 Gbaud/s.

Table 2 provides a summary of our results and a comparison with other graphene EA modulators. Our graphene EA modulator exhibits notable advantages compared to other graphene EA modulators listed in Table 2in terms of modulation depth and modulation efficiency. Compared to devices utilizing mechanical exfoliation of graphene (Ref 76), our platform offers the capability to transfer large-area CVD graphene onto substrates enabling efficient and cost-effective production of graphene-based devices (Supplementary Note 1). The bandwidth of the device can be further optimized by increasing the quality of the oxide layer, which could also lead to improvements in the overall device performances.

**Table 2. Performances of the graphene EA modulators**

| Device structure | Insertion loss (dB/μm) | Modulation depth (dB) | Voltage (V) | Modulation efficiency (dB/V) | L (μm) | Normalized modulation efficiency (dB·V$^{-1}$·μm$^{-1}$) | Bandwidth (GHz) | Data rate (Gb/s) | Power consumption (fJ/bit) |
|---|---|---|---|---|---|---|---|---|---|
| DLG(a) | 0.025 | 7.1 | 7 | 1.3 | 180 | 0.007 | 26.8 | 48 | 200 |
| DLG MZI(a) | 0.1 | 13.3 | 3 | 5.3 | 200 | 0.0265 | 7.4 | 32 | 200 |
| DLG [29] | 0.092 | 3 | 9 | 1.2 | 120 | 0.01 | 29 | 50 | N/A |
| DLG [45] | 0.13 | 4 | 12 | 2.2 | 60 | 0.037 | 39 | 40 | 160 |
| SLG [27] | 0.076 | 5 | 8 | 1.5 | 50 | 0.03 | 5.9 | 10 | 350 |
| SLG [25] | N/A | 3.6 | 4.5 | 0.95 | 40 | 0.0237 | 1 | N/A | N/A |

(a) This work;

## 3. Conclusion

In summary, we proposed high-performance graphene integrated photonic platform based on the gold-assist transfer. Thanks to the higher quality transfer method, our devices clearly display performances beyond state-of-the-art. In particular, we demonstrated a graphene TO micro-ring modulator with a compact active area of 7.54 μm$^2$, with wavelength tuning efficiency of ~0.037 nm·mW$^{-1}$, heating efficiency of ~67.4 K·μm$^3$·mW$^{-1}$ and power consumption per FSR of ~81 mW. The real-time optical measurement of this device demonstrated operation at 50 kHz with rise and fall response times of 4.8 and 5.5 μs, respectively. Next, by replacing the oxide layer, we demonstrated a graphene EA modulator with a 3 dB bandwidth up to 26.8 GHz and a high data rate reaching 48 Gb/s. Furthermore, we demonstrate a graphene MZI modulator features a high modulation depth up to 13.3 dB at bias voltage of 5 V and an impressive modulation efficiency up to 5.3 dB/V corresponding to a normalization modulation efficiency of 0.027 dB·V$^{-1}$·μm$^{-1}$. The realized devices display high efficiency, high speed, small footprint, low energy consumption and CMOS-compatibility, and they thus hold great potential for large-scale production of micro-scale integrated opto-electronic devices, with foreseeable applications in the fields of neural networks, ultrafast data communication, biosensing, and LIDAR.

## 4. Methods

### 4.1 Device Fabrication

The waveguides were fabricated using a photo-lithography process by a CMOS foundry. A 10 nm of BCB is spin-coated on top to planarize and promote adhesion of the surface. Commercially available SLG (BGI) grown on copper foil via CVD are transferred using the gold-assisted transfer method. The gold supporting layer is deposited on the surface of a SLG sample and then it is placed into a solution of 1 g of ammonium persulfate (1 g diluted in 40 mL of DI water) to etch the copper substrate. Once the copper etching is complete, the gold-SLG film is transferred in a beaker with DI water to wash off residual etchant. After several repetitions of cleaning, lift the gold-SLG film onto target substrate and wait for drying. Subsequently, graphene was lithographically patterned followed by wet etching with acid solution and oxygen plasma etching respectively to define active channels. After a dielectric layer is deposited on top by atomic layer deposition, we repeated the graphene-transfer process to construct the SLG/dielectric/SLG structure.

*4.2 Experimental Setup*

The transmission spectra under different voltages were measured with the measurement system shown in Supplementary Note 6. Two probes were utilized to establish contact with the electrodes, allowing the application of an electrical signal from a source meter to the device under test (DUT) through gold electrodes. The input light in the TE mode, generated from a tunable laser source, was carefully polarization controlled and launched into the chip by a grating coupler subsequently. The spectral response was collected by coupling the output light back to optical fiber and connected to an optical power meter.

**Funding.** National Key Research and Development-Program of China (2019YFB2205303), National Natural Science Foundation of China (62205054, 62375051), Natural Science Foundation of Jiangsu Province (BK20210207), and Fundamental Research Funds for the Central Universities. G.S. acknowledges funding from the Deutsche Forschungsgemeinschaft (DFG, German Research Foundation) through the Collaborative Research Center (CRC) 1375 "NOA", project C4.

**Author Contributions.** X. X. Wu† and Z. Y. Cao† contributed equally to this work. J. W. conceived the idea for the project. J.W., G.S. and E.L. supervised the work. X. W., and T. Z. designed the devices. Y. W., Z. C. and Z. L. fabricated the devices. X. W., S. D., E. L., O.G. and G. S. performed the simulations. X. W. Y. X. and L. L. measured devices. X. W. did the data analysis and drafted the manuscript. All authors have contributed to the manuscript and gave approval of the final version.

**Disclosures.** The authors declare no competing financial interests.

**Acknowledgements.** For the purpose of open access, the authors have applied a Creative Commons Attribution (CC BY) license to the article.

**Data availability.** Data underlying the results presented in this paper are not publicly available at this time but may be obtained from the authors upon reasonable request.

**Supplemental document.** See Supplement 1 for supporting content.

## REFERENCES

1. S. Y. Siew, B. Li, F. Gao, H. Y. Zheng, W. Zhang, P. Guo, S. W. Xie, A. Song, B. Dong, L. W. Luo, C. Li, X. Luo, and G.-Q. Lo, "Review of Silicon Photonics Technology and Platform Development," J. Light. Technol. **39**, 4374–4389 (2021).
2. R. Soref, "The Past, Present, and Future of Silicon Photonics," IEEE J. Sel. Top. Quantum Electron. **12**, 1678–1687 (2006).
3. Q. Xu, S. Manipatruni, B. Schmidt, J. Shakya, and M. Lipson, "125 Gbit/s carrier-injection-based silicon micro-ring silicon modulators," Opt. Express **15**, 430 (2007).
4. Y. Cao, V. Fatemi, S. Fang, K. Watanabe, T. Taniguchi, E. Kaxiras, and P. Jarillo-Herrero, "Unconventional superconductivity in magic-angle graphene superlattices," Nature **556**, 43–50 (2018).
5. M. Romagnoli, V. Sorianello, M. Midrio, F. H. L. Koppens, C. Huyghebaert, D. Neumaier, P. Galli, W. Templ, A. D'Errico, and A. C. Ferrari, "Graphene-based integrated photonics for next-generation datacom and telecom," Nat. Rev. Mater. **3**, 392–414 (2018).
6. N. Youngblood, Y. Anugrah, R. Ma, S. J. Koester, and M. Li, "Multifunctional Graphene Optical Modulator and Photodetector Integrated on Silicon Waveguides," Nano Lett. **14**, 2741–2746 (2014).
7. E. Heidari, H. Dalir, F. M. Koushyar, B. M. Nouri, C. Patil, M. Miscuglio, D. Akinwande, and V. J. Sorger, "Integrated ultra-high-performance graphene optical modulator," Nanophotonics **11**, 4011–4016 (2022).
8. R. Amin, J. K. George, H. Wang, R. Maiti, Z. Ma, H. Dalir, J. B. Khurgin, and V. J. Sorger, "An ITO–graphene heterojunction integrated absorption modulator on Si-photonics for neuromorphic nonlinear activation," APL Photonics **6**, 120801 (2021).
9. M. A. Giambra, V. Mišeikis, S. Pezzini, S. Marconi, A. Montanaro, F. Fabbri, V. Sorianello, A. C. Ferrari, C. Coletti, and M. Romagnoli, "Wafer-Scale Integration of Graphene-Based Photonic Devices," ACS Nano **15**, 3171–3187 (2021).
10. B. S. Lee, A. P. Freitas, A. Gil-Molina, E. Shim, Y. Zhu, J. Hone, and M. Lipson, "Scalable graphene platform for Tbits/s data transmission, arXiv 2020," arXiv preprint arXiv:2011.08832 (2020).
11. C. Wu, S. Brems, D. Yudistira, D. Cott, A. Milenin, K. Vandersmissen, A. Maestre, A. Centeno, A. Zurutuza, J. Van Campenhout, C. Huyghebaert, D. Van Thourhout, and M. Pantouvaki, "Wafer-Scale Integration of Single Layer Graphene Electro-Absorption Modulators in a 300 mm CMOS Pilot Line," Laser Photonics Rev. **17**, 2200789 (2023).
12. X. Gan, R.-J. Shiue, Y. Gao, I. Meric, T. F. Heinz, K. Shepard, J. Hone, S. Assefa, and D. Englund, "Chip-integrated ultrafast graphene photodetector with high responsivity," Nat. Photonics **7**, 883–887 (2013).
13. T.-H. Han, Y. Lee, M.-R. Choi, S.-H. Woo, S.-H. Bae, B. H. Hong, J.-H. Ahn, and T.-W. Lee, "Extremely efficient flexible organic light-emitting diodes with modified graphene anode," Nat. Photonics **6**, 105–110 (2012).

14. X. Du, I. Skachko, A. Barker, and E. Y. Andrei, "Approaching ballistic transport in suspended graphene," Nat. Nanotechnol. **3**, 491–495 (2008).
15. M. Massicotte, G. Soavi, A. Principi, and K.-J. Tielrooij, "Hot carriers in graphene – fundamentals and applications," Nanoscale **13**, 8376–8411 (2021).
16. A. A. Balandin, S. Ghosh, W. Bao, I. Calizo, D. Teweldebrhan, F. Miao, and C. N. Lau, "Superior Thermal Conductivity of Single-Layer Graphene," Nano Lett. **8**, 902–907 (2008).
17. O. Dogadov, C. Trovatello, B. Yao, G. Soavi, and G. Cerullo, "Parametric Nonlinear Optics with Layered Materials and Related Heterostructures," Laser Photonics Rev. **16**, 2100726 (2022).
18. H. Li, Y. Anugrah, S. J. Koester, and M. Li, "Optical absorption in graphene integrated on silicon waveguides," Appl. Phys. Lett. **101**, 111110 (2012).
19. L. Yu, D. Dai, and S. He, "Graphene-based transparent flexible heat conductor for thermally tuning nanophotonic integrated devices," Appl. Phys. Lett. **105**, 251104 (2014).
20. Y. Sun, Y. Cao, Y. Yi, L. Tian, Y. Zheng, J. Zheng, F. Wang, and D. Zhang, "A low-power consumption MZI thermal optical switch with a graphene-assisted heating layer and air trench," RSC Adv. **7**, 39922–39927 (2017).
21. S. Nakamura, K. Sekiya, S. Matano, Y. Shimura, Y. Nakade, K. Nakagawa, Y. Monnai, and H. Maki, "High-Speed and On-Chip Optical Switch Based on a Graphene Microheater," ACS Nano **16**, 2690–2698 (2022).
22. D. Schall, M. Mohsin, A. A. Sagade, M. Otto, B. Chmielak, S. Suckow, A. L. Giesecke, D. Neumaier, and H. Kurz, "Infrared transparent graphene heater for silicon photonic integrated circuits," Opt. Express **24**, 7871 (2016).
23. S. Yan, X. Zhu, L. H. Frandsen, S. Xiao, N. A. Mortensen, J. Dong, and Y. Ding, "Slow-light-enhanced energy efficiency for graphene microheaters on silicon photonic crystal waveguides," Nat. Commun. **8**, 14411 (2017).
24. Z. Xu, C. Qiu, Y. Yang, Q. Zhu, X. Jiang, Y. Zhang, W. Gao, and Y. Su, "Ultra-compact tunable silicon nanobeam cavity with an energy-efficient graphene micro-heater," Opt. Express **25**, 19479 (2017).
25. M. Liu, X. Yin, E. Ulin-Avila, B. Geng, T. Zentgraf, L. Ju, F. Wang, and X. Zhang, "A graphene-based broadband optical modulator," Nature **474**, 64–67 (2011).
26. A. Phatak, Z. Cheng, C. Qin, and K. Goda, "Design of electro-optic modulators based on graphene-on-silicon slot waveguides," Opt. Lett. **41**, 2501 (2016).
27. Y. Hu, M. Pantouvaki, J. Van Campenhout, S. Brems, I. Asselberghs, C. Huyghebaert, P. Absil, and D. Van Thourhout, "Broadband 10 Gb/s operation of graphene electro-absorption modulator on silicon: Broadband 10 Gb/s operation of graphene electro-absorption modulator on silicon," Laser Photonics Rev. **10**, 307–316 (2016).
28. Z. Cheng, X. Zhu, M. Galili, L. H. Frandsen, H. Hu, S. Xiao, J. Dong, Y. Ding, L. K. Oxenløwe, and X. Zhang, "Double-layer graphene on photonic crystal waveguide electro-absorption modulator with 12 GHz bandwidth," Nanophotonics **9**, 2377–2385 (2020).
29. M. A. Giambra, V. Sorianello, V. Miseikis, S. Marconi, A. Montanaro, P. Galli, S. Pezzini, C. Coletti, and M. Romagnoli, "High-speed double layer graphene electro-absorption modulator on SOI waveguide," Opt. Express **27**, 20145 (2019).
30. D. Ansell, I. P. Radko, Z. Han, F. J. Rodriguez, S. I. Bozhevolnyi, and A. N. Grigorenko, "Hybrid graphene plasmonic waveguide modulators," Nat. Commun. **6**, 8846 (2015).
31. C. Qiu, H. Zeng, and Y. Su, "Recent progress in graphene-based optical modulators on silicon photonics platform," Natl. Sci. Open **1**, 20220031 (2022).
32. C. Luan, Y. Liu, Y. Ding, and H. Hu, "2-um high-speed graphene electro-optic modulator based on silicon slot microring resonator," in *Conference on Lasers and Electro-Optics (CLEO)*, JTh2F.34.
33. H. Dalir, Y. Xia, Y. Wang, and X. Zhang, "Athermal Broadband Graphene Optical Modulator with 35 GHz Speed," ACS Photonics **3**, 1564–1568 (2016).
34. Y. Liao, G. Feng, H. Zhou, J. Mo, H. J. Sun, and S. H. Zhou, "Ultra-Broadband All-Optical Graphene Modulator," IEEE Photonics Technol. Lett. **30**, 661–664 (2018).
35. P. Dong, W. Qian, H. Liang, R. Shafiiha, N.-N. Feng, D. Feng, X. Zheng, A. V. Krishnamoorthy, and M. Asghari, "Low power and compact reconfigurable multiplexing devices based on silicon microring resonators," Opt. Express **18**, 9852 (2010).
36. X. Liu, P. Ying, X. Zhong, J. Xu, Y. Han, S. Yu, and X. Cai, "Highly efficient thermo-optic tunable micro-ring resonator based on an LNOI platform," Opt. Lett. **45**, 6318 (2020).
37. M. Onodera, S. Masubuchi, R. Moriya, and T. Machida, "Assembly of van der Waals heterostructures: exfoliation, searching, and stacking of 2D materials," Jpn. J. Appl. Phys. **59**, 010101 (2020).
38. A. Montanaro, W. Wei, D. De Fazio, U. Sassi, G. Soavi, P. Aversa, A. C. Ferrari, H. Happy, P. Legagneux, and E. Pallecchi, "Optoelectronic mixing with high-frequency graphene transistors," Nat. Commun. **12**, 2728 (2021).
39. H. Van Ngoc, Y. Qian, S. K. Han, and D. J. Kang, "PMMA-Etching-Free Transfer of Wafer-scale Chemical Vapor Deposition Two-dimensional Atomic Crystal by a Water Soluble Polyvinyl Alcohol Polymer Method," Sci. Rep. **6**, 33096 (2016).
40. X. Gao, L. Zheng, F. Luo, J. Qian, J. Wang, M. Yan, W. Wang, Q. Wu, J. Tang, Y. Cao, C. Tan, J. Tang, M. Zhu, Y. Wang, Y. Li, L. Sun, G. Gao, J. Yin, L. Lin, Z. Liu, S. Qin, and H. Peng, "Integrated wafer-scale ultra-flat graphene by gradient surface energy modulation," Nat. Commun. **13**, 5410 (2022).
41. Y.-M. Seo, W. Jang, T. Gu, H.-J. Seok, S. Han, B. L. Choi, H.-K. Kim, H. Chae, J. Kang, and D. Whang, "Defect-Free Mechanical Graphene Transfer Using *n*- Doping Adhesive Gel Buffer," ACS Nano **15**, 11276–11284 (2021).

42. X. Liang, B. A. Sperling, I. Calizo, G. Cheng, C. A. Hacker, Q. Zhang, Y. Obeng, K. Yan, H. Peng, Q. Li, X. Zhu, H. Yuan, A. R. Hight Walker, Z. Liu, L. Peng, and C. A. Richter, "Toward Clean and Crackless Transfer of Graphene," ACS Nano **5**, 9144–9153 (2011).
43. Y. Wu, X. Zou, M. Sun, Z. Cao, X. Wang, S. Huo, J. Zhou, Y. Yang, X. Yu, Y. Kong, G. Yu, L. Liao, and T. Chen, "200 GHz Maximum Oscillation Frequency in CVD Graphene Radio Frequency Transistors," ACS Appl. Mater. Interfaces **8**, 25645–25649 (2016).
44. R. S. Sundaram, M. Steiner, H.-Y. Chiu, M. Engel, A. A. Bol, R. Krupke, M. Burghard, K. Kern, and P. Avouris, "The Graphene–Gold Interface and Its Implications for Nanoelectronics," Nano Lett. **11**, 3833–3837 (2011).
45. H. Agarwal, B. Terrés, L. Orsini, A. Montanaro, V. Sorianello, M. Pantouvaki, K. Watanabe, T. Taniguchi, D. V. Thourhout, M. Romagnoli, and F. H. L. Koppens, "2D-3D integration of hexagonal boron nitride and a high-κ dielectric for ultrafast graphene-based electro-absorption modulators," Nat. Commun. **12**, 1070 (2021).